\documentclass[article,twocolumn,superscriptaddress,amsmath,amssymb]{revtex4}
\usepackage{cancel}
\usepackage{maybemath}
\usepackage{xcolor}
\usepackage{graphicx}\newcommand{\dd}{\mathrm{d}}
\newcommand{\ii}{\mathrm{i}}
\newcommand{\ee}{\mathrm{e}}
\newcommand{\calM}{\mathcal{M}}
\newcommand{\calS}{\mathcal{S}}
\newcommand{\calT}{\mathcal{T}}
\newcommand{\thr}{\mathrm{th}}
\newcommand{\vectornotation}[1]{\ensuremath{\mathbf{#1}}}
\begin{document}

\title{Do SN 1987A data yield the three neutrino masses?}

\author{Robert Ehrlich}
\affiliation{George Mason University, Fairfax, VA 22030, United States} 
\email{rehrlich@gmu.edu}
\date{\today}
\begin{abstract} Finding the neutrino mass is critical because it breaks the Standard Model of particle physics in which neutrinos are massless and it provides the key to understanding the evolution of the universe.  Yet currently only upper limits are known for the neutrino masses based on cosmological constraints and direct mass experiments.  This paper explores the unlikely possibility that SN 1987A might provide actual values for the three neutrino masses and not just upper limits, a possibility first suggested by Ramanath Cowsik in 1988.  Cowsik's result depends on the neutrino emissions from SN 1987A being near-simultaneous, i.e., within a time interval $\Delta t<1 s.$  Having such a brief burst, however, is contradicted by virtually all supernova neutrino emission models that include a $\Delta t>10 s$ cooling phase.  Here it is explained why those neutrino emission models may be mistaken, and why the three neutrino masses suggested by the SN 1987A data might be correct, even though they are larger than the upper limits implied from cosmology and direct mass experiments.  
\end{abstract}
\maketitle.
\section{Introduction}
About $10^{58}$ neutrinos were emitted from the dying star Sanduleak 69 202 when it became the first supernova seen in 1987 (hence the name SN 1987A).  Of those $10^{58}$ neutrinos only a few dozen were captured by Earth's detectors. Yet, thousands of papers have been published about SN 1987A, and its well-studied nature is not surprising given that it is the only supernova in or close to our galaxy since neutrino detectors first existed. The earliest articles on the neutrino mass obtained from SN 1987A data, with one exception,\citep{Cowsik1988} set only upper limits on its value.~\citep{Arnett1987, Burrows1988, Fang1987}.   The upper limits restriction is based on the almost universal belief that the values of the three neutrino mass states are simply too small to use the SN 1987A data to find them.  

Thus, it is usually assumed that the spread in neutrino emission times for SN 1987A is large enough to disguise any tiny differences in neutrino travel times from which the neutrino mass could be found,  Here we make no such assumption, and instead use the SN 1987A data themselves to see what they reveal about whether the masses can be found.  As a result, the author rediscovered what Ramanath Cowsik had found in 1988 about the neutrino masses being possibly revealed in the SN 1987A data.~\citep{Ehrlich2011}  As we shall see, the data suggests $m^2$ values for the three neutrino masses, all of which are much larger than thought possible, and one of which is a tachyon, i.e., $m^2<0.$~\citep{Bilaniuk1962}.  While this is not the first time the author has made this claim, the paper has a number of new elements, including (1)  an improved statistical analysis, (2) a demonstration that finding individual neutrino masses is really possible, (3) an extensive critique of the standard neutrino emission models of SN 1987A, and (4) explanations of why each of the constraints seemingly forbidding $m^2<0$ neutrinos can be circumvented. 

\subsection{Are $m^2<0,$ and $v>c$ Tachyons ``unphysical"?} Many theorists have regarded quantum fields giving rise to physical tachyons as being either impossible or unlikely in light of the many theoretical difficulties they create.  However, unobservable $m^2<0$ particles seem to be acceptable to theorists.  Thus, for most theorists the idea that symmetry-breaking in the interaction of Higgs fields with massless gauge fields leads to the production of unobservable tachyons is not objectionable. The difficulties with ``physical" tachyons having an $m^2<0$ that can be observed include their energy spectrum being unbounded from below, their frame-dependent and unstable vacuum state, and their noncovariant commutation rules.  Nevertheless, there are also theorists who are accepting of physical tachyons, and recently Paczos et al.~\citep{Paczos2024, Paczos2025}  have shown that a covariant framework exists which allows for the proper quantization of tachyon fields which eliminates all such ``unphysical" issues.  Charles Schwartz is another theorist who has successfully challenged many the reasons why tachyons are considered unphysical, and shown how tachyons can account for some cosmological observations.~\citep{Schwartz2018, Schwartz2022}.  The interest here, however, is not on tachyon neutrino theory, but primarily on data which either supports or contradicts the possibility that neutrinos are tachyons.

\subsection{Do negative searches make tachyons less likely?}
Most reports of  tachyons, including the 2011 OPERA experiment initial $v>c$ result,~\citep{Brumfiel2012} have later been shown to be incorrect apart from one-time sightings.~\citep{Ehrlich2022} Many physicists therefore believe that the tachyon hypothesis can be dismissed, or at least relegated to the highly implausible category.  A more reasonable conclusion is that as long as not a single neutrino has ever been measured to have either $m^2>0$ or $v<c,$ the tachyonic neutrino hypothesis remains viable.  A decline in belief in physical tachyons is perfectly understandable as a psychological reaction to negative results or even worse, false positive results like OPERA.   However, negative and false positive results do not suggest tachyons do not exist, only that the $m^2$ for neutrinos is so close to zero, that we cannot yet tell if $m^2>0$ or $m^2<0$. 

You might think a negative hypothetical result $m^2=0.0\pm 0.1 eV^2/c^4$ makes tachyonic neutrinos more unlikely than a previous result  $m^2=0.0\pm 10.0 eV^2/c^4$ because the allowed negative range in $m^2$ has been reduced a hundredfold.  However, a Bayesian would realize that conclusion is false, because the allowed positive range has also been reduced by the same hundredfold amount.  Neutrinos of course are the only known particles that might be tachyons since all others have been observed travelling at speeds $v<c,$ and tachyons if they exist must always be superluminal.  One area in which such specific predictions of tachyonic neutrinos have been made is that of direct neutrino mass experiments.

\subsection{Direct mass experiments}
To date most experiments to measure the mass of the electron neutrino (or antineutrino) have done so by fitting the shape of the beta decay spectrum of tritium ${^3H}\rightarrow {^3He} + e + \bar{\nu_e}$ very close to its endpoint.  These experiments actually measure an ``effective mass" defined in terms of the three states having masses $m_j$ and the PMNS mixing matrix $U_{i,j},$ applicable to the three flavor states as

\begin{equation}
m_{\nu,i}^2=\Sigma |U_{i,j}|^2m_j^2
\end{equation}

where for the electron, muon and tau neutrinos $i=1,2,3,$ respectively.  Some experiments in the past have reported negative values for $m_\nu^2,$ for the electron neutrino, but these were due to systematic effects, and not a true $m_\nu^2<0.$  The KATRIN experiment is the latest and most accurate experiment to measure $m_\nu^2$ and it has corrected for many of those earlier systematic effects, especially molecular final states.~\citep{Schneidewind2024,Drexlin2026}  As of 2024 KATRIN has found only an upper limit to the mass, $m_\nu<0.45 eV/c^2$ and a best fit value is $m^2=-0.14^{+0.13}_{-0.15} eV^2.$~\citep{KATRIN2024}  This result, of course, leaves the true sign of $m^2$ ambiguous given the uncertainty.  It is interesting, however, that this KATRIN value for $m_\nu^2$ reported in 2024 is consistent with the value of the  tachyonic neutrino mass that the author claimed in a 2015 paper based on six unrelated observations, i.e., $m^2=-0.11\pm0.02 eV^2.$ -- see Table 1 in ref. ~\citep{Ehrlich2015}  

It is claimed by KATRIN collaborators that direct mass experiments are ``model-independent" and they make no assumptions (aside from energy conservation).~\citep{Drexlin2013, KATRIN2022, Parno2025}  However, there is one assumption always made, yet not acknowledged. Namely, it is assumed that if the electron neutrino effective mass is close to zero, then the three masses $m_j$ contributing to it in Eq. 1 must also be close to zero.  This assumption is important because when one makes it, only a single mass need be used to fit the spectrum. However, the falsity of this assumption is easy to see, because if one of the three masses were to satisfy $m_j^2<0,$ Eq. 1 would allow the effective masses for $\nu_e, \nu_\mu$ and $\nu_\tau$ to all be very close to zero, even when the individual $m_j$ are not.  In fact, it will be shown using published neutrino data for SN 1987A that the three $m_j,  j=1,2,3$  comprising the electron neutrino may have surprisingly large values.  In fact, the three masses from SN 1987A are grossly in conflict with the widely accepted normal or inverted hierarchy, for which the $m^2_j$ separations are no greater than about $\Delta m^2=0.00243eV^2.$  

\subsection{Fit to the $\beta$-spectrum using one effective mass}
If the widely separated large neutrino masses here inferred from SN 1987A data are correct, one might imagine that either (a) good fits to a single effective mass could not possibly be observed in a high precision experiment like KATRIN, or (b) such good fits should show some hint of the three masses $m_j$.  Specifically, one might expect to observe either kinks in the differential spectrum or excess counts in the {\emph {integrated}} spectrum near its endpoint in the vicinity of $E=E_0-m_j$ for $j=1,2,3$  As a matter of fact, when the initial KATRIN results were published in 2019 the author did report spotting just such a hint of excess positive residuals (number of sigma) in the data based on a spectrum fit using a single effective mass.~\citep{Ehrlich2019a}.  

In an experiment as complex as KATRIN it is useful to divide the data taking into measurement campaigns to accommodate necessary equipment upgrades, and allow researchers to optimize experimental conditions between campaigns. 
Thus, after the initial data release researchers implemented several hardware changes and operational adjustments to mitigate and eliminate backgrounds that could cause artificial distortions in the beta-decay spectrum.  Hopefully, the spectral distortions so eliminated would not include those indicative of new physics that might produce anomalies when fitting the data to a single effective mass instead of three separate masses.  Of the five measurement campaigns reported in 2024, one (KNM5) had almost half the total number of counts (16.7 million out of 36 million).  It also is the most stable of the five data sets~\citep{KATRIN2024}.  The KNM5 campaign data showed a similar hint of excess positive residuals as was reported by the author following the 2019 initial data release -- see Fig.1, particularly the the energy interval about 2 to 10 eV from the spectrum endpoint.~\citep{KATRIN2024}.

\begin{figure}
\centerline{\includegraphics[angle=0,width=1.10\linewidth]{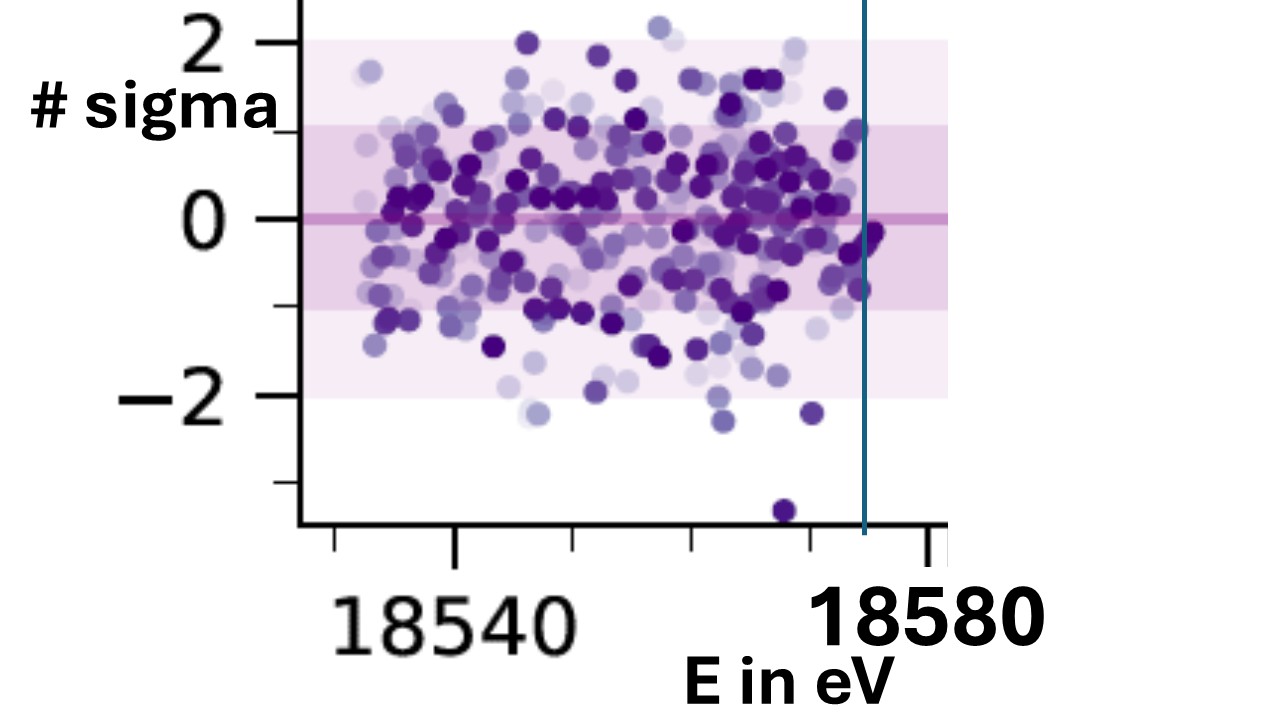}}
\caption{Residuals plot for spectrum fit using a single effective mass for measurement campaign KNM5, with the spectrum endpoint at $E_0=18574$ identified by the author as a vertical line.  The plot has is a cropped version from that in ref. \citep{KATRIN2024} (license CC BY 4.0).}
\end{figure}

\subsection{Supernova SN 1987A neutrinos} 
Three neutrino detectors operated at the time supernova SN1987A occurred at 7:35 UT on Feb 23, 1987.  Those detectors (Kamiokande II, IMB, and Baksan) recorded ``bursts" or seconds-long groupings of 12, 8 and 5 events respectively at approximately the same time, and they recorded the neutrino visible energy, incoming direction and arrival times.  The three detectors were not perfectly synchronized, with a time difference between them being a matter of some debate.  One recent analysis has concluded that there exist substantial differences between the time of the first arriving neutrino in each detector.~\citep{Bozza2026}.  According to this analysis, the first event in Kamiokande II arrived $6.4 \pm 60 sec$ before that in IMB, and the first event in Baksan arrived $t=30^{+2}_{-60}$ sec after that in IMB -- see Table I in ref.~\citep{Bozza2026}.  Clearly, with such large uncertainties in the claimed detector desynchronizations, it would be useless to combine the data if our goal is to find a more reliable temporal distribution.  We therefore simply adopt the usual convention of setting the arrival time of each detector's first event to $t=0$.  As one estimate of how much uncertanty in time this assumption introduces, we note that the average gap beween the first and second event in the three detectors is about 0.3 seconds.

The arriving antineutrinos were detected based on inverse beta decay
$\bar{\nu_e}+p \rightarrow n + e^+$ so that the antineutrino energy was actually 1.3 MeV higher than the recorded visible $(e^+)$ energy.  Those three neutrino bursts occurred about three hours before the star emitting them suddenly brightened.  This late light arrival, of course does not imply that neutrinos are superluminal.
Rather, the three-hour gap between the neutrino burst and the later arrival of visible light, aligns with the small interaction between neutrinos with matter.  Thus, photons but not neutrinos are delayed by interactions with the supernova ejecta.  

\subsection {The LSD neutrino burst} In addition to the three detectors at the time of SN 1987A there was a fourth small 
detector known as LSD which was located under Mont Blanc.  This detector was in fact better shielded from background radiation than the other three, and had a lower energy threshold, so in these two respects it was more sensitive.  The LSD detector reported a burst of five neutrinos which occurred 484 min before the other three detectors' recorded their bursts.  This early burst is therefore considered by most physicists to be a background fluctuation unrelated to SN 1987A because of its five hour early arrival. If the LSD burst were genuine it could constitute evidence that those neutrinos were tachyons, although other explanations also exist, including having SN1987A ``bang" twice~\citep{Galeotti2016} and other ones as well.~\citep{Agafonova2018}  Nevertheless, even raising the possibility of tachyons probably was enough to dismiss the existence of the LSD neutrino burst as unrelated to SN 1987A in the minds of most physicists.  Moreover, there are five other reasons to be suspicious about the LSD burst besides the  5 hr early arrival, starting with those neutrinos not being able to reach Earth.  

According to Andrew Cohen and Sheldon Glashow superluminal neutrinos would decay via $\nu\rightarrow\nu e^+e^-.$~\citep{Cohen2010}  Many such decays repeated over any appreciable distance would lower the neutrino energy into nothingness.  Now, the neutrinos detected in LSD had a degree of superluminality $\delta=\Delta t/T=484 min/168,000 yr= 5.5\times 10^{-9}.$  In this case, the threshold energy for this decay process $E_{th}=2m_e/\sqrt{\delta}=13.8 GeV,$ which is far greater than the energies of the LSD neutrinos, so the decay process proposed by Cohen-Glashow would not occur.  Another argument against the LSD burst being due to superluminal neutrinos is that it is simply a statistical fluctuation of background.  Having five neutrinos observed during a 7 second time interval in LSD by chance would represent a very significant departure from background (roughly one neutrino every 80 seconds), and that background had been steady for months. In fact, the probability of a chance fluctuation giving rise to the LSD signal is extremely low, less than $1.4 \times 10^{–6}.$~\citep{Galeotti2016}  Many observers also reject the LSD burst because no such early burst was seen in the three other detectors.  However, the LSD neutrinos were all very low energy, and the other detectors could not have seen an LSD-like burst at that early time because of their higher energy thresholds.  Thus, for example, the largest detector, Kamiokande II, had an energy threshold of 7 MeV, well above the energies of four of the five LSD neutrinos constituting the burst  Another suspicious fact is that the main t=0 burst at 7:35 UT was not seen in LSD.  However, that can also be explained because the small LSD detector did observe two neutrinos that time (not enough to qualify as a burst).  Two final major problems with the LSD burst being associated with SN 1987A are (1) the very low energies of all the 5 neutrinos comprising it, and (2) their nearly equal energies or the virtual monochromaticity of the burst.  In  what follows we accept that the LSD neutrino burst was associated with SN 1987A, because all five objections to it are answerable, but we address the monochromaticity problem later.

\section {Finding $m^2$ of SN 1987A neutrinos}
The possibility of using the SN 1987A data to find the three neutrino masses may seem implausible, but we shall see that it can be done if the neutrinos were emitted from the star close enough to simultaneously.  Let us first see why for exactly simultaneous neutrino emissions the neutrino arrival time relative to a photon directly determines their travel time (relative to light), and how using the measured neutrino energy $E,$ and arrival time t we can compute the mass of the neutrinos.   In what follows $T=168,000$ years is the time photons required to reach us from SN 1987A.

If a photon's speed en route to Earth is $c=D/T$ then a neutrino's speed would be $v=D/(t+T)\approx c(1-t/T).$ Here t is the time delay of a neutrino relative to a photon.  But since there can also be such a delay if a neutrino is emitted later than other neutrinos, so the definition of t  only makes sense if almost all neutrinos were emitted during a very small time window.  For the neutrino speed we have  $v/c =1-t/T.$ and also as an excellent approximation $v/c=\sqrt{1-m^2c^4/E^2}\approx 1-m^2c^4/2E^2$.  Setting the two versions of $v/c$ equal 
\begin{equation}
v/c=1-t/T=1-m^2c^4/2E^2
\end{equation}
yields
\begin{equation}
 \frac{1}{E^2}=\big(\frac{2}{Tm^2c^4}\big)t\equiv Mt
\end{equation}

Eq. 3 implies that if a number of neutrinos with the same mass $m^2c^4$ were present and they had different arrival times t and energies E, then on a plot of $1/E^2$ versus $t,$ the data point for each neutrino will be found to lie on or near a straight line through the origin of slope M given by

\begin{equation}
M = \frac{2}{Tm^2c^4}
\end{equation}

Furthermore, if the SN 1987A data show such a clustering about a straight line, from the best-fit slope M we can find the common mass of all neutrinos on or near the line as:
\begin{equation}
m^2c^4=\frac{2}{TM}
\end{equation}
Obviously steeper sloped lines yield smaller masses, a photon lies on a vertical line, and $m^2<0$ neutrinos lie on a negatively sloped line through the origin lying in the $t<0$ region.

To the extent that neutrinos all started out from SN 1987A at nearly the same instant, then since there are known to be three (and only three) non-sterile neutrino masses, every neutrino should lie on or close to one of three straight lines passing through the origin.  Any neutrinos that do not do so must either be a background event unrelated to SN 1987A, or else the result of non-simultaneous emissions.  It is however possible that the data points will all be found to be associated with only one or two straight lines rather than three lines if two or three of the lines (or neutrino masses) are too close to distinguish from one another. 

\section{Fitting 30 data points to 3 lines}
Thirty neutrinos is not a very large number, but the distinctiveness of a pattern counts for far more than the sheer number of events.  Thus, the $\Omega^-$ particle which led to quarks was confirmed in 1964 at Brookhaven National Laboratory by observing just one single``Gold Standard" event.~\citep{Barnes1964}  In the present case, the pattern found when the SN 1987A data is plotted (see Fig.2) is also very distinctive since every one of the neutrinos from SN 1987A lies on or next to one of three straight lines through the origin as predicted, and only one data point very near the origin is within an error bar width of two lines.  The least squares fit of 30 data points to three lines having a common $t=0$ origin has three free parameters, $M_j$ the slopes of the three straight lines in Fig. 2 yielding 27 degrees of freedom.  The parameter H, the size of the constant horizontal error bar on each data point in Fig. 2 is also adjusted -- see Fig. 3 -- where it can be seen that the value $H=0.5$ seconds or larger gives a reasonable fit.  It is reassuring that this value of $H$ is within a factor of two of the earlier uncertainty inferred from the average time (0.3 s) between the first and second event in each detector. Obviously, the uncertainty of each neutrino arrival $t$ is known to far greater precision than $H=0.5$ sec, which measures the uncertainty caused by setting the burst arrival equal to that of the first event in each detector.  Since each data point has both x and y error bars, we can then convert the x error into an "effective" y error by multiplying the x uncertainty by the slope of the line, and combine it with the y uncertainty in quadrature.  If a data point lies close to two straight lines, we simply choose the line that gives the smaller contribution to chi square.  

\subsection{The best fit and a 3 + 3 model}
The best fit of 30 data points $(t,1/E^2),$ one for each neutrino to one of three straight lines that pass through the origin is shown in Fig. 2.  Error bars in both horizontal and vertical are included in the fit.  The three best fit masses $m_j$ are obtained from the three best fit slopes according to Eq. 5.  Note the different scales used for $t>0$ and $t<0$  The five LSD data points at $t=-29 ksec$ have been slightly separated from each other for clarity. The good fit shown in Fig 2 has a chi square of 28.6, and $p=38\%.$ This fit was for a horizontal error bar of $\pm 0.5$ seconds on all data points.  For a fuller description of how the $\chi^2$ fit probability depends on the size of the horizontal error bar see Fig. 3. We have seen that exactly simultaneous neutrino emissions requires  the data must lie on one of three lines through the origin in a $1/E^2$ versus t plot.  Therefore the goodness of fit using $\pm0.5$ sec horizontal error bars shows  the points will continue to lie on or near these lines if the time window of most neutrino emissions is no greater than about one second.  

\begin{figure}
\centerline{\includegraphics[angle=0,width=1.0\linewidth]{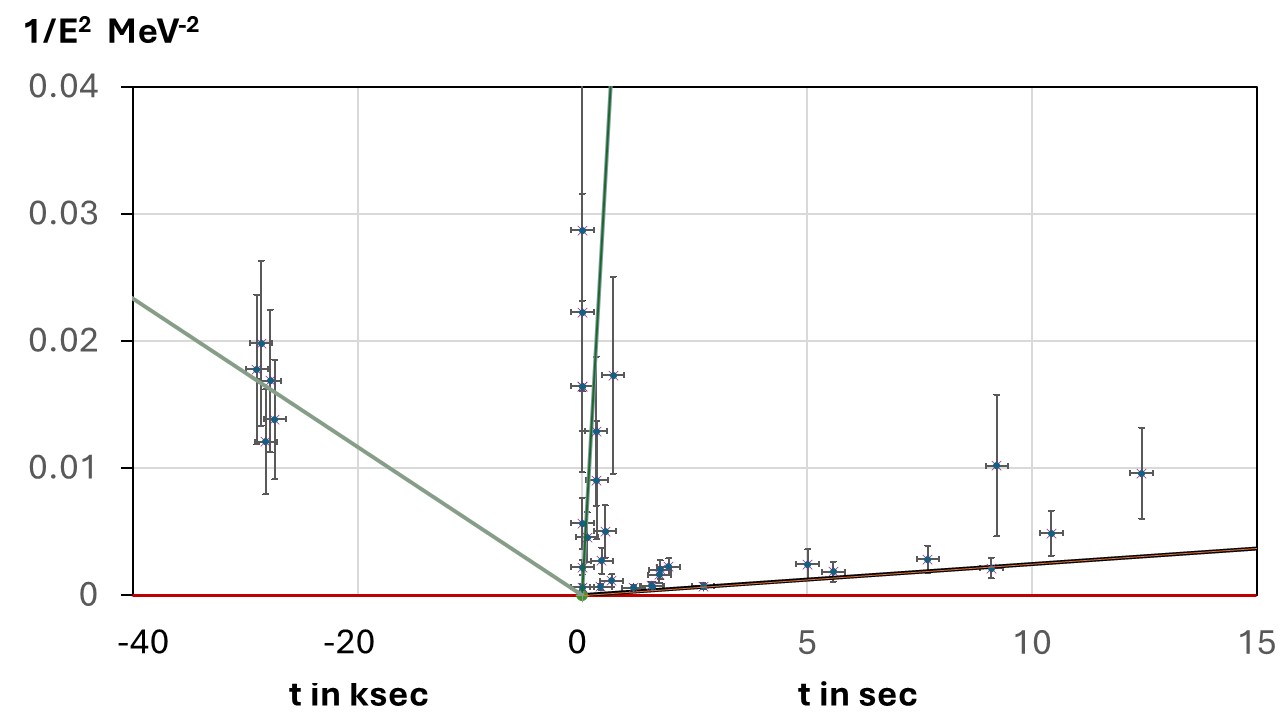}}
\caption{Best fit of 30 data points $(t,1/E^2),$ one for each neutrino to one of three straight lines that pass through the origin.  The very large vertical error bars on those points having small energy E are a consequence of error propagation: $\Delta(1/E^2)=2\Delta E/E^3.$}
\end{figure}

Based on Eq. 5 the three masses corresponding to the slopes of the three straight lines in Fig. 2 are $m_1=2.0 eV/c^2$ and $m_2=22.4 eV/c^2$ for the positive sloped lines and $m_3^2=-680,000eV^2/c^4$ for the $m^2<0$ line.  The author first discovered this three line fit to the  SN 1987A data in 2013, when he interpreted the result (with significantly different numerical values for $m_1$ and $m_3$) in terms of a 3 + 3 model, i.e., three pairs of nearly degenerate active-sterile neutrinos -- see Fig.4.)~\citep{Ehrlich2013,Ehrlich2019}  Of course, the usual model with three sterile neutrinos is based on the seesaw mechnism, where the sterile neutrinos are much heavier than the active ones, but here their masses are nearly the same, which is not disallowed by any basic physics principles. 

Given such large masses as the SN 1987A data suggests, this model needs active-sterile mixing and not active-active in order to create the $\Delta m^2$ observed in atmospheric and solar neutrino oscillations.  Having three sterile neutrinos as shown in Fig. 4 is attractive for various reasons.  First, having three of them would restore the symmetry between quarks and leptons, since then every left quark and lepton has a right counterpart.  In addition, to comply with theories of leptogenesis and dark matter, there must be at least 3 flavors of sterile neutrinos (if they exist).~\citep{Masahiro2016}  Finally, giving further theoretical support to having three sterile neutrinos, Goncalves and colleages have shown that three sterile neutrinos are needed in conformal Majoran models to cancel anomalies seen in various oscillation experiements.~\citep{Goncalves2026}  

The main differences from the author's previously published work on the good fit to the SN 1987A data is a change in the value of $m_2$ because of the past omission of horizontal error bars.  There was also regretably a drastically incorrect value reported for $m_3^2.$   We are not unaware of the interesting approximate numerical coincidence of the value of $m_2=2.0 eV/c^2$ and the $2.7 eV/c^2$ sterile neutrino mass reported by the Neutrino-4 collaboration, as well as its {\emph {non}}-observation by KATRIN,~\citep{Serebrov2024,KATRIN2025} a matter discussed at length in section~\ref{Dog}.  

\subsection{Corroboration of the two $m^2>0$ masses}\label{sterile}
The existence of sterile neutrinos is highly controversial, and there are probably more negative search results than positive ones.  Thus, neither MicroBooNE, nor IceCube, nor KATRIN nor DayaBay report any evidence for them.~\citep{Micro2025, Aartsen2020, KATRIN2025, An2024}.  Nevertheless, an omission in these and other experiments is that most have not sought evidence for the presence of multiple sterile neutrinos simultaneously.  There are two possible remedies for this omission, the simplest being to search for two or three sterile neutrinos in the same fit, and another one being to find a place where nature causes a spatial separation between light and heavy sterile neutrinos.  

To date the only search seeking to verify the pair of $m^2>0$ masses of sterile neutrinos suggested by the SN 1987A data has been carried out by M. H. Chan and the author.~\citep{Chan2014}  Sterile neutrinos are a natural possibility for the identity of dark matter in the galaxy and clusters of galaxies.  Essentially, Chan and Ehrlich first derived an equation of state for the sterile neutrinos dependent on their mass, and then separately found the dark matter radial profile in both the Milky Way and four groups of galaxies.  Finally, the sterile neutrino mass was varied to find the value that best fit the observed dark matter profile in each case. The results of these fits showed that the two $m^2>0$ values from the 3-mass SN 1987A fit would indeed fit the galaxy clusters (lighter mass) and the Milky Way (heavier mass). This result offers some confirmation of the masses deduced from the SN 1987A three mass fit.  However, the good fits found are less impressive than they might have been, because with a revised value for $m_1=2 eV$ -- half the original value -- the fit to the four galaxy clusters is not nearly as good as it was.

\begin{figure}
\centerline{\includegraphics[angle=0,width=1.1\linewidth]{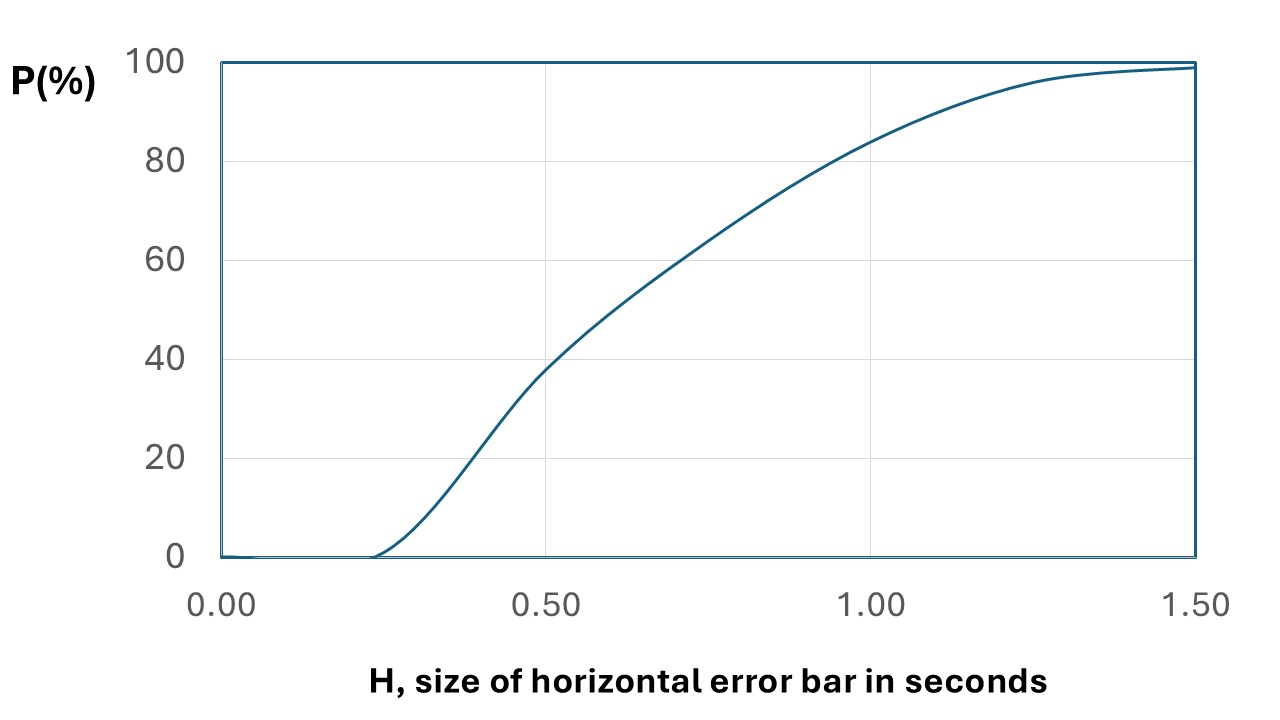}}
\caption{The probability from the least squares fit to three straight lines through the origin versus the size of the horizontal error bars, assumed to be the same for all 30 data points.  Note that there is no ``best" fit to three straight lines unless one first fixes the size of the horizontal error bars because for $H>1.5$ sec error bars the fit probability $p\approx100\%$}
\end{figure}
\begin{figure}
\centerline{\includegraphics[angle=0,width=1.0\linewidth]{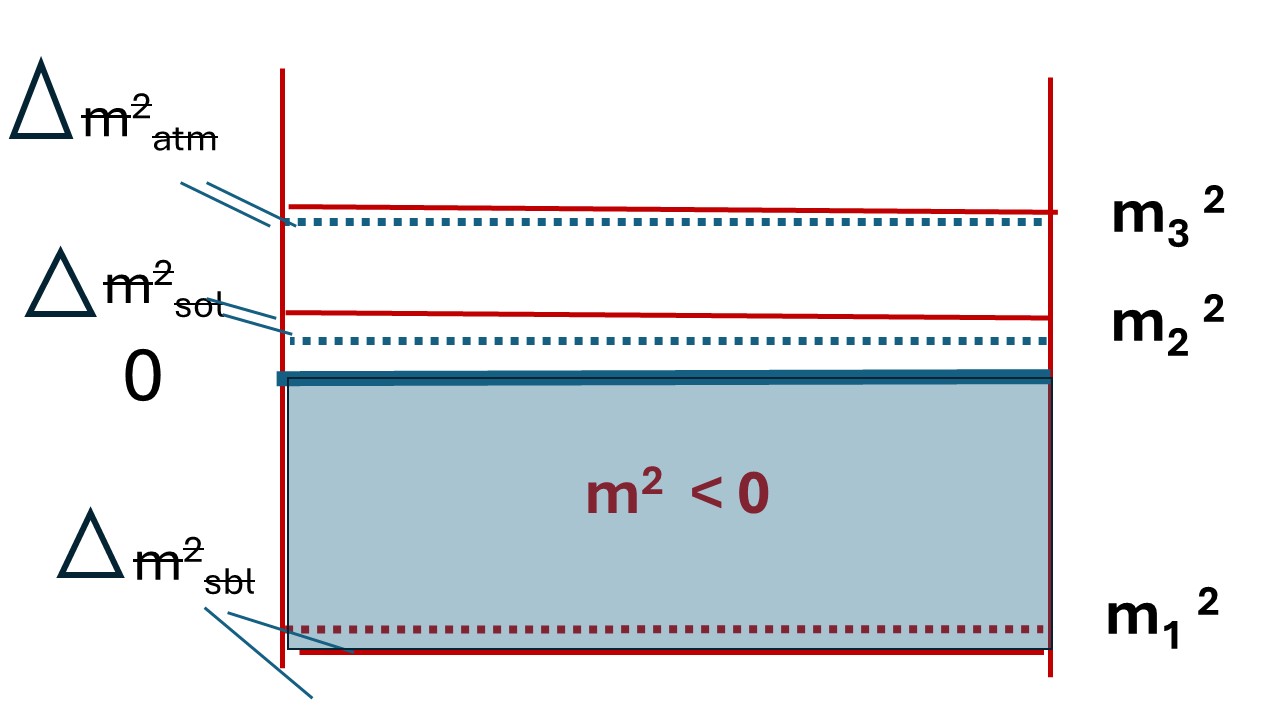}}
\caption{$3+3$ neutrino mass squared diagram postulated by the author in 2013.  There are 3 active-sterile doublets split by three $\Delta m^2$ for solar, atmospheric and short baseline (sbl) oscillations. The lowest mass doublet has $m^2<0.$}
\end{figure}

\section{Discussion}
In this section we discuss a wide range of issues that need to be addressed before we can have any confidence that the excellent fit in fig.2 to three straight lines (three neutrino masses), including one with $m^2<0$ is not just a mirage or background fluctuation.  

\subsection{Comparison of $\nu$ emission models for SN 1987A\label{nocooling}}
The main problem with the author's model is that it assumes near simultaneous neutrino emissions from SN 1987A, and this requirement is at variance with standard neutrino emissions models for supernovae.  Those models usually stipulate two emission phases: an intense brief accretion phase lasting less than a few tenths of a second followed by a $10 +$ seconds long cooling phase.~\citep{Burrows1986}  In contrast, in the author's model there are also two temporal groupings of neutrinos, but they are the result of a single brief burst being bifurcated on arrival at Earth by having two neutrino travel time distributions due to two different $m_\nu^2>0$ masses.  

In this section we show why the conventional model is unproven, and possibly wrong.  More specifically, we show why the existence of a $10+$ seconds long large amplitude cooling phase for SN 1987A is not firmly established.  This does not mean we claim there is no cooling phase for neutrino emissions, only that any such phase might have (a) very small amplitude, or (b) duration much different from 10 seconds (much longer or shorter).  If one of these is true, the cooling phase can be assumed to have merged with the accretion phase, in effect making for a single phase model, in which the neutrino luminocity can be approximated by a delta function.   

The use of a delta function should give a good fit to the data if the actual emission luminocity were described by a function having a full width at half max (FWHM) of one second, as we found when obtaining a good fit using a horizontal error bar of $H=\pm0.5 s.$  A very small amplitude cooling phase is really the same as a very long duration because in that case the enormous heat of the proto-neutron star could be shed over a longer time at a much slower rate, and there will be a negligible number of cooling phase neutrinos in the 10 second region being analyzed.  On the other hand, the other possibility of a very short duration cooling phase probably would require physics beyond the standard model, such as:
\begin{enumerate}
\item {\emph{Fast-flavor conversions,}} in which the density of neutrinos is so immense that they begin to scatter off each other rather than just standard matter.  This process can drastically increase the cooling layer's efficiency, and shrink the cooling timeline.
\item {\emph{Exotic particle cooling,}} in which hypothetical particles, like sterile neutrinos or axions, which could act as a ``shortcut" for energy escape.
\item {\emph{Neutrino-devouring dark matter}} that interacts with neutrinos can "devour" them within the core, resulting in excessive cooling and a shortened burst
\end{enumerate}

A typical standard neutrino emission model has been created by Bozza et al., who have calculated the neutrino luminosity from SN 1987A having a $10+$ second cooling phase following an accretion phase.~\citep{Bozza2025} Their resulting computed luminosity is determined by a least squares fit to the SN 1987A data, and is shown in Fig. 5.  Although the figure seems to show a much smaller integrated luminocity for $t>1$ second than for $t<1$ second, the two are of similar size due to the change in scale at $t=1$ sec.  This Bozza et al. fit to the data has 9 free parameters, unlike the author's fit, which involves zero free parameters.  It is zero free parameters because it uses a delta function to describe neutrino emission time dependence at its source, but of course {\emph {not}} for the observed time dependence at Earth.   If two supernova neutrino emission models both give good fits to the SN 1987A neutrino data, the only data yet available, we may ask which model is more likely to be closest to reality, simplicity being an important criteria.  On simplicity grounds, it should not be difficult choosing between a zero free parameter fit versus one that has nine free parameters, since with that many the model is capable of fitting virtually any shape emissions time dependence.    

Further favoring a one-phase model for SN 1987A is the fact that its cooling was peculiar and unexpected because it originated from a blue supergiant, rather than a red supergiant.  This fact would lead to faster adiabatic cooling and a fainter initial object, which has challenged traditional models having an extended $10+ sec$ cooling phase.  The cooling phase continues to be puzzling, and it has prompted a resort to highly exotic cooling mechanisms like ``axions from strange matter."~\citep{Cavan-Piton2024}  These difficulties with the cooling phase data for SN 1987A were explored at length in Fiorillo et al.~\citep{Fiorillo2023} who note that: ``While our models show compatibility with the events detected during the first seconds, [they are] in conflict with the late-event bunches in Kam-II and Baksan after 8–9 s, which are also difficult to explain by background."  

The real justification for having a $10 +$ sec cooling phase in the Bozza and other conventional models is because that is what SN 1987A data seems to show, however, that is {\emph{only}} if the neutrino masses are small enough to ignore the differences in their energy-dependent travel times. Very small neutrino masses are simply assumed in the Bozza model based on cosmological data and direct mass experiments, not the SN 1987A data in itself. Consistency with direct mass experiments can be achieved if one mass has $m_j^2<0,$ as previously explained, and consistency with cosmological data can be achieved under the same condition, as explained in a future section.

\begin{figure}[t]
\centerline{\includegraphics[angle=0,width=1.10\linewidth]{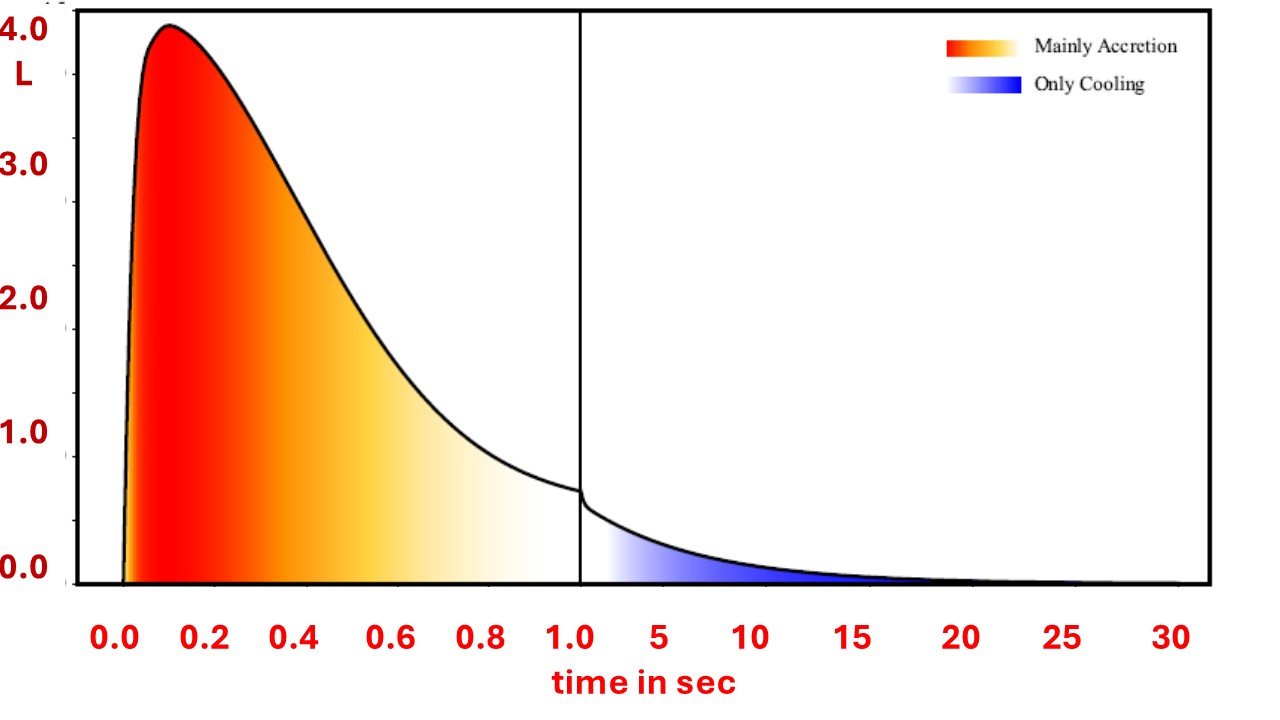}}
\caption{Computed temporal electron antineutrino luminosity L in units of $10^{53}erg/s$ for SN 1987A neutrinos from Bozza et al.~\citep{Bozza2025} The accretion phase luminosity is in red, and the cooling phase is in blue.  Note the time scale change at $t=1$ seconds.  Figure is used under creative commons license (CC BY 4.0)}
\end{figure}

\begin{figure}[t]
\centerline{\includegraphics[angle=0,width=1.\linewidth]{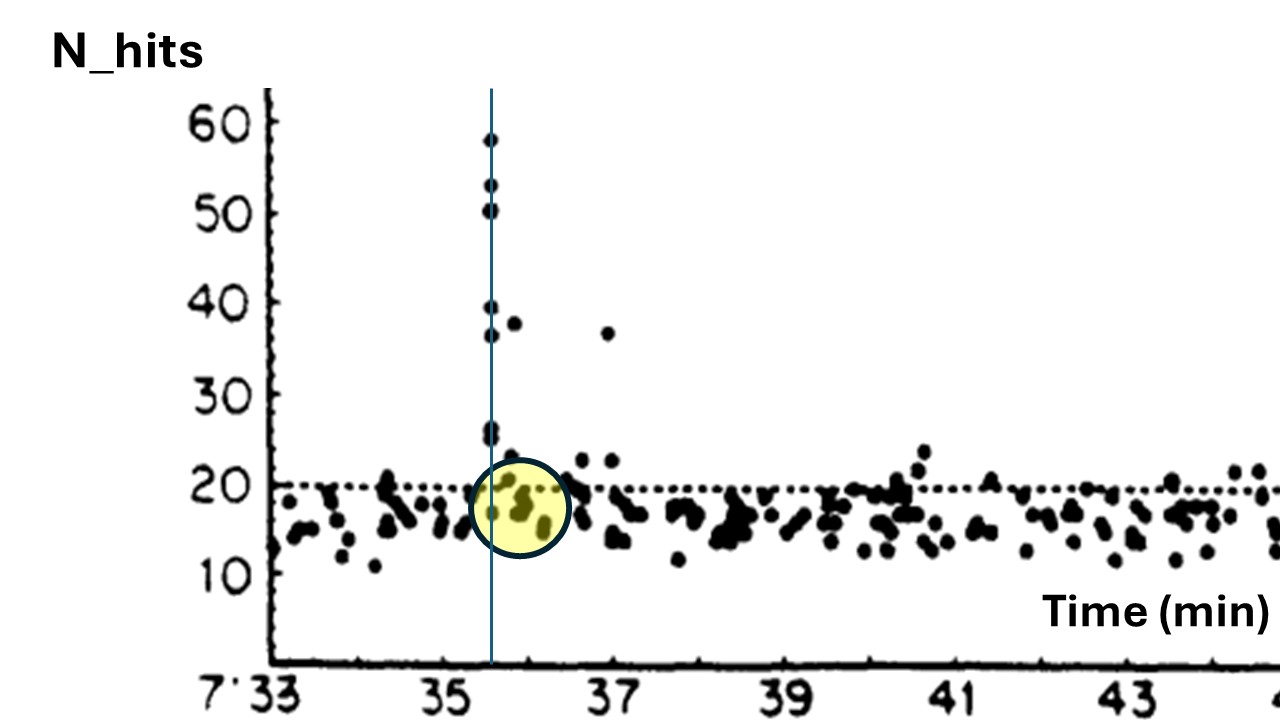}}
\caption{The Kamiokande II data, i.e., the number of hits versus time in the minutes before and after SN 1987A from Hirata et al~\citep{Hirata1988}.  The number of hits is the number of photomultipliers activated, and is a measure of the neutrino energy.  The vertical line at the time 7:35:35 shows the time of the main burst.  The small circle added by the author highlights four extra data points identified by Bozza et al., based on lowering the threshold. Three of these four points are located right at the circle center. license (CC BY 4.0)}
\end{figure}
\subsection{Finding four extra Kamiokande II events?}
The most controversial aspect of the Bozza et al. analysis involves their choice of threshold for distinguishing Kamiokande II neutrinos associated with SN 1987A from background.  There were 12 events listed in the 1988 Kamiokande II paper as being associated with SN 1987A burst in ref.~\citep{Hirata1988}, and all but one of these events have $N_{hits}>20$ or $E>7.5 MeV$.  Bozza et al has, however, ``discovered" an additional four events, based on setting a much lower energy threshold $E> 4.6 MeV$ or $N_{hits}=12.$  Three of those events comprise the irregular blob at the circle center in Fig. 6 well below the dotted horizontal line.  Bozza etal. believes the lowered threshold is defensible because one of the original 12 Kamiokande II events has that low energy, but that event occurred less than one second of $t=0.$  More importantly, the lowered threshold clearly risks misidentifying four bckground events as being SN 1987A-associated.  In addition, adding these four questionable events significantly increases the cooling phase amplitude, since the four events occur $t>10 s$, more than doubling the number of events originally present there.  Moreover, if the four circled events really are part of a SN 1987A cooling phase, there would seem to be every reason for Bozza to then include the much less questionable three events having well above 20 hits that are located above the label $t=37$ min in Fig. 6.  The sprinkling of very high energy neutrinos continues well beyond the end of the graph at $t=43$ min. In fact, the 17 minute time interval after the main burst in the Kamiokande II data in ref. ~\citep{Hirata1988}) reveals 25 neutrinos above 20 hits, which is almost twice the average number in the other seven 17 min time intervals Kamiokande II displayed -- see Fig. 7.  The excess number of neutrinos having more than 20 hits in the time interval after the main burst might indicate a cooling phase of duration more like $+10$ minutes than $+10$ seconds.  Cooling phases as long as 10 minutes were not considered in Bozza.~\citep{Bozza2025}  For such a long cooling phase the chances of are very small of even one cooling phase neutrino appearing in a plot like Fig. 2 extending only up to $t=15 s.$ That means that the delta function approximation for neutrino emissions would be excellent with such a long cooling phase.

\begin{figure}[t]
\centerline{\includegraphics[angle=0,width=1.\linewidth]{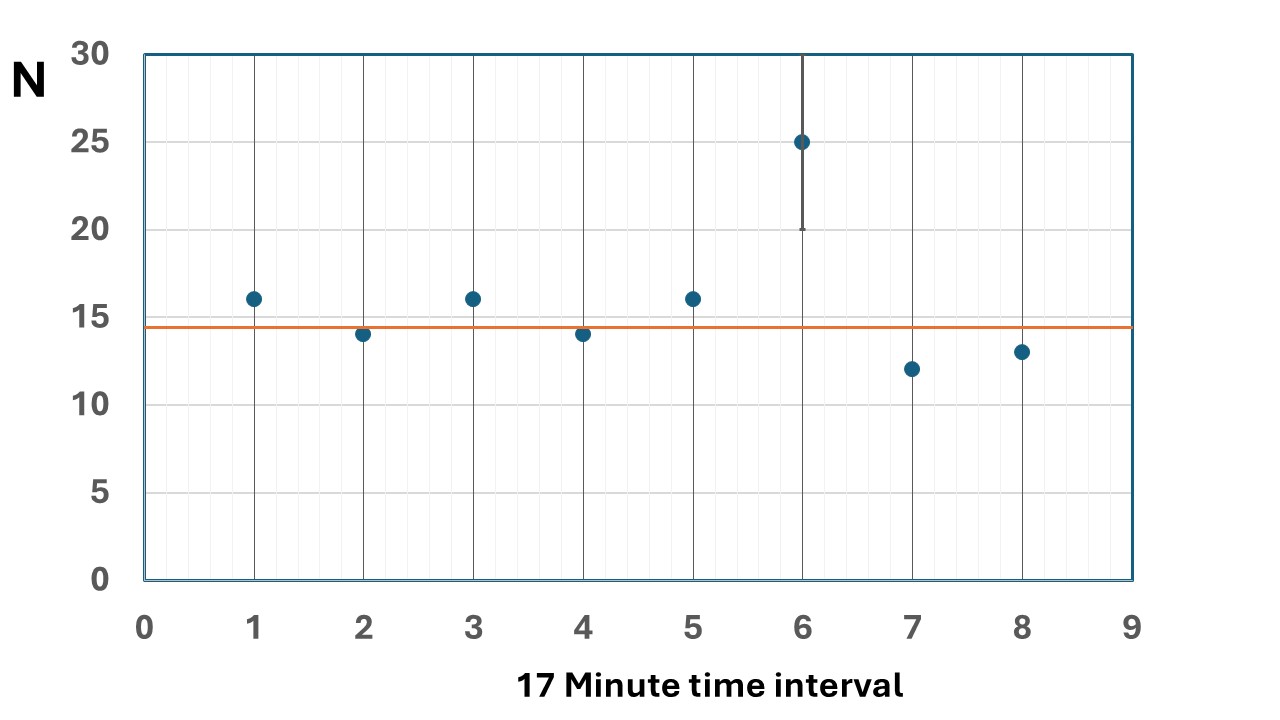}}
\caption{The Kamiokande II data from ref.~\citep{Hirata1988}, showing the number of events $N$ having $N_{hits}>20$ in each of eight non-contiguous 17-min-long time intervals arranged in sequential order.  The main burst is near the beginning of time interval 5, and events within 12 seconds of the burst are not included in the count. The data are suggestive of a $\Delta t >10$ minute cooling phase based on the excess counts in time interval 6.}
\end{figure}

\subsection{Resolving the ambiguity between models}
While the Bozza model is clearly more complex than the author's model, both models give good fits to the SN 1987A data and neither model can claim to be the correct one -- they merely explain the $10+$ second observed spread in neutrino arrival times in different ways.  The Bozza model attributes the spread to a cooling phase lasting $10+$ seconds, and the author's model explains it by a spread in neutrino travel times due to their different masses, with a much shorter or more likely longer cooling phase.  Which explanation is correct will not be known until the arrival of a new galactic supernova.  Once a new supernova appears both the Bozza approach of ignoring differences in neutrino travel times and the author's approach of assuming near-simultaneous (delta function) emissions should both once again give good fits to the neutrino data, but with one important difference.   

Let's say a new supernova occurs at half the distance of SN 1987A, i.e., $T=\frac{1}{2}(168,000)$ years.  If the author's model is correct, the three lines found in the fit of $1/E^2$ vs $t$ will by Eq.5  have twice their slope, but the same masses of course.  In contrast, the cooling phase luminosity graph in the Bozza model will unphysically drop off with time twice as fast.  In contrast, if the Bozza model is the correct one, the fit to the data will look just like that in Fig. 5, and so will the the $1/E^2$ vs $t$ fit look just like Fig. 2, but given the dependence on $T$ in Eq. 5, that implies masses twice as large as found from SN 1987A.  Clearly, the correct explanation should become completely clear once a new supernova appears, especially given the far greater size of today's neutrino detectors than those in 1987A.  Until then, there are two major advantages of the author's zero free parameter model over a nine free parameter model like Bozza's, namely the author's model naturally accommodates the $t<0$ LSD neutrino burst, but of course, it does so at the cost of invoking the existence of $m^2<0$ neutrinos.  The second advantage is that only the author's model is testable in a variety of non-supernova contexts.

\subsection{Issues with the LSD burst}
Initially, the author did reject the LSD data.  This rejection was not because it provided evidence for an $m^2<0$ neutrino, but because the idea of monochromatic neutrinos from a supernova seemed preposterous.  However since this issue and most of the other reasons many people have for rejecting the LSD data can be dealt with, there is no valid reason not to keep LSD data, especially because given three active neutrinos, the mandated number of straight lines about which clustering of data points in Fig. 2 should occur is three.  The burden of proof here should not be why not exclude this LSD burst, but rather why not keep it -- and there really is no compelling reason to ignore it.

\subsubsection{Why are the LSD data so bunched in time?}
If the LSD burst of 5 neutrinos  is genuine, it may look strange in Fig. 2 that while the other data points lie all along their respective lines the five LSD neutrinos are all bunched together in time at $t=-484$ min or -29 ksec. The three straight line result would be more believable perhaps if the five neutrinos yielded data points at five widely spaced points on the negatively sloped line.  But recall that the $t<0$ data have a kilosecond time scale.  If the five LSD data points were spread out along the LSD line and not bunched in time they never could have been distinguished from background because researchers were looking for a concentration of neutrino events in a seconds-long time interval, i.e., a burst.  The fact that the 5 LSD neutrinos were recorded at -29 ksec does, of course, not mean they ``went backward in time," only that they travelled faster than light, assuming they started at $t=0$ along with all the other three neutrino bursts. 
\subsubsection{Why do the LSD neutrinos have the same E?}
The apparent near-constant energy of the LSD burst remains mysterious even to the physicists who accept its existence, but it is natural if the 5 neutrinos (and all others) must lie on one of three straight lines. Moreover, there is a theoretical basis for the monochromaticity of the LSD burst, as shown in several of the author's publications that have provided evidence for it.~\citep{Ehrlich2021} That evidence was in the form of a proposed model for an 8 MeV neutrino line from a supernova, based on the proposed $X17$ or $Z'$ particle (of mass 17 MeV) discovered by the Krasznahorkay group.~\citep{Krasznahorkay2024}  In this model the monochromatic neutrinos would be created from annhilating cold dark matter X particles in the reaction 
\begin{equation}
X\bar{X}\rightarrow Z'\rightarrow \nu \bar{\nu}
\end{equation}
The evidence for the model was provided using the only other leptonic decay of $Z'$
\begin{equation}
X\bar{X}\rightarrow Z'\rightarrow e^+e^-
\end{equation}
which accounts for the observed spectrum of MeV $\gamma-$rays from the galactic center.~\citep{Ehrlich2021}

\subsubsection{Alternate explanations for the LSD neutrinos?}
There are other explanations of the LSD burst, besides tachyons but they each have problems.  Thus, perhaps the most likely alternate possibility is that a supernova can bang more than once.  In fact, Galeotti and Pizzella have even claimed there was a third bang-- not in the LSD detector but in Kamiokande II at a time about 20 min after the main burst.~\citep{Galeotti2007}  Supernovae undergoing multiple bangs is a known phenomenon (known as Pulsational pair-instability supernovae), which can occur for very massive stars according to Woosley.~\citep{Woosley2017}  However, the mass of SN 1987A was too small to experience the pair instability and in any case the phenomenon are not known to produce detectable neutrino bursts. 

There are other explanations of the LSD burst.

\begin{itemize}
\item If it was a statistical fluctuation it was a really big one $p=1.4\times 10^{-6}.$ In fact among the LSD signal candidates for the detection of a neutrino burst selected over 14 years of operation, it has the lowest background imitation probability.
\item If it was a genuine, low-energy pulse of neutrinos (e.g., from neutron capture on iron slabs around the detector), simulations show that their number is negligible.~\citep{Manukovskiy2022}. 
\item If it was due to 8 MeV dark matter as suggested in ref.~\citep{Ehrlich2021} this would still create the same LSD burst, so it is not really an alternative, but rather an explanation of the LSD burst.
\item The only other explanation known to this author for the LSD burst besides tachyons is the Froggatt-Nielsen proposal of ``dark matter pearl-size balls" inside the stellar core~\citep{Froggatt2015}.
\end{itemize}

\subsection{The $\sum m_{\nu }$ constraint from cosmology} 
A conflict exists between the three large neutrino masses from the fit to SN 1987A data and cosmological data implying small neutrino masses. How might such a conflict be avoided?  In the weak-field approximations of general relativity, tachyons would apparently experience a repulsive gravitational force from normal matter, rather than the attraction normally observed.\citep{Raychaudhuri1974}  To the extent that subsequent structure formation in the early universe depends on the value of $\sum m_{\nu },$ then if one or more neutrinos are tachyons, it then follows that their masses must enter the mass sum as negative  not imaginary values.  Otherwise, we would not possibly have cancellation of the effects of attractive and repulsive forces, that is between or structure building and destroying effects.
With a negative value for the tachyonic mass, not only might that term in $\Sigma m_k$ be negative, but so might the entire sum, which may indeed be the case in the real universe.  In fact, the current best value derived from analysis of the cosmic microwave background (CMB) and baryon acoustic oscillations (BAO) from the Dark Energy Spectroscopic Instrument (DESI) favors a negative value $\sum m_{\nu }=-0.16\pm 0.09 eV,$ which is likely to persist even as more data are accumulated.~\citep{Green2025}  

Few cosmologists believe this result implies true negative masses, instead attributing it to a tension between different data sets or a sign of new physics, although Jamie Farnes has proposed a unifying theory of dark energy and dark matter involving negative neutrino masses and matter creation within a modified $\Lambda CDM$ framework.~\citep{Farnes2018} Farnes paper states that it was motivated based upon a statement by Albert Einstein, who had written that the cosmological constant required that ``empty space takes the role of gravitating negative masses which are distributed all over the interstellar space"  Even with negative mass neutrinos, however, there is still an obvious conflict between the current small limit on the mass sum and the three large masses found from the SN 1987A data.  One way to avoid this conflict would be if the mass sum were over flavor state effective masses and not mass state masses, as is usually assumed.  With the sum being over flavor state masses, they could all be very small, if one mass state is a tachyon with $m^2<0$, as explained earlier.  Of course, flavor states only have ``effective" masses, but that need not mean sums over flavor state masses are meaningless.  In fact it is common to use a summation over effective masses to allow for the possibility of negative energy density.~\citep{Elbers2025}

One way to justify the neutrino mass sum as being over the flavor state effective masses has been provided by Lesgourgues and Pastor,~\citep{Lesgourgues2012} who note note that ``cosmology is at first order sensitive to the total neutrino mass {\emph {``if all states have the same number density,"}}which we refer to as the Lesgourges-Pastor criterion.  The connection between this criterion and doing the mass sum over flavor states becomes obvious if we consider the mass sum after sone neutrino producing process in the early universe
when it was about 1 second old.  At that time the two dominant processes capable of producing neutrinos were $Z^0$ decay or $e^+e^-$ collisions into neutrino antineutrino pairs, and both reactions would produce equal numbers of neutrinos for each flavor.  For example, $Z^0$ decays would result in equal $6.7\%$ branching ratios for each flavor, thus creating an equal number density $N$ of each flavor.  Since it is the flavor masses not the mass eigenstates that had the same number density, thus based on the Lesgourgues-Pastor criterion the relevant ``total mass" immediately following their birth is:  
\begin{equation}
N\sum m_\nu = Nm_{\nu,e}+Nm_{\nu,\mu}+Nm_{\nu,\tau}
\end{equation}
Of course, the neutrinos born in flavor states soon become an incoherent sum of mass states as they propagate through space  However, that fact would not change their global mass density computed from the sum over previously equally populated flavor states.   

\subsection{A ``missing" $2.70 eV/c^2$ sterile neutrino?}
In 2025 KATRIN reported finding no evidence for sterile neutrinos based on the observed shape of the tritium beta decay spectrum.~\citep{KATRIN2025}.  It conducted this search by fitting its data to a sum of two spectra: that for a zero mass neutrino plus the spectrum for a sterile neutrino of mass $m_{4}$, which was assumed to take on various values and various weights $\alpha=sin^2 2\theta.$   At each $(m_4,\alpha)$ grid point, the $\chi^2$ function is minimized with respect to three free parameters.  By examining the excluded regions of the plane based on a high $\chi^2$ KATRIN concluded there was no evidence for sterile neutrinos at a high confidence level in their data over the entire $(m_4,\alpha)$ grid.  This negative result was in conflict with the Neutrino-4 group that had previously found evidence for a $2.70 eV/c^2$ sterile neutrino in their oscillation experiment.  Moreover, the conflict of KATRIN with the Neutrino-4 experiment was not a small one, since the data of the GALLEX, SAGE, and BEST experiments confirms the parameters of neutrino oscillations $m_4=2.70 eV/c^2$ and $\sin^22\theta_{14} = 0.36)$ and the confidence level of the Neutrino-4 sterile neutrinos combined with the data of the GALLEX, SAGE, and BEST experiments is now $5.8\sigma.$~\citep{Serebrov2024}  

\subsection{Why KATRIN failed to see sterile neutrinos?}\label{Dog}
It is possible that KATRIN failed to detect any sterile neutrinos because they do not exist, but here we consider a more interesting possibility.  Neutrino - 4 and the other three experiments that saw the $2.70 eV/c^2$ neutrino were doing a short baseline disappearance oscillation experiment, which observed the disappearance probability $P(\bar{\nu_e}\rightarrow \bar{\nu_e}).$  This type of experiment would work just as well if the disappearing neutrinos were disappearing into sterile or active neutrinos.  However, KATRIN's search was specifically looking for sterile neutrinos and it would not have worked for active ones as it was conducted.  If the main claim of this paper is correct, and if the $2.0 eV/c^2$ neutrino is an active neutrino, then KATRIN's failure to see it is understandable because their search method was not appropriate.    They should have instead fit their data using a combined spectrum $R_C$ for the three adjustible masses $m_{\nu,j}$ and weights, $\alpha_j,$ that is:
\begin{equation}
R_{C}=\sum\alpha_jR_\beta(E,E_0,m^2_{\nu,j})+R_{Bkgd}
\end{equation} 
where $R_{Bkgd}$ is the constant background rate, $R_\beta$ is the usual beta decay formula for a given $m^2$, E is the electron energy and $E_0$ is the fitted endpoint energy for tritium beta decay.  They could then see if a good fit can be found in order to test the validity of the main claim of this paper (that the three mass fit of SN 1987A data yields the three active neutrino masses), KATRIN could see where on the $m_1$ and $m_2$ plane a maximum likelihood is found.  A ``best possible outcome" would be a plot like that in Fig. 8 showing a maximum likelihood found near the pair of $m^2>0$ masses found from the SN 1987A data.  For the functional form of the contribution to the spectrum for the $m^2<0$ third mass, KATRIN could just use the usual beta decay formula, and simply substitute a negative $m^2$ in it.   While Rembielinski, Caban and Ciborowski have shown that their tachyonic neutrino model results in a more complicated formula -- see Eqs. 93-94 in ref.~\citep{Rembielinski2021}.  However, as shown in their Fig. 1, the simple approach gives a result that is virtually indistinguishable from their full description.\\

\subsection{Multiple sterile neutrinos in oscillation data?}
The need for doing searches for multiple sterile neutrinos in a single search has already been noted in section \ref{sterile}.  How would one  go about seeking evidence for both $m_1=2.0 eV$ and $m_2=22.4 eV$ in oscillation data, bearing in mind that the oscillation wavelength for $m_2$ would be on the order of 100 times shorter than for the $m_1$ sterile neutrinos? Obviously, such different wavelengths would require both a near and far detector as in the DUNE, NOvA, T2K, SBN, and MINOS+ experiments.  One approach to find the two oscillations would involve setting up a grid of values for the two masses $m_1, m_2$ and in each case use the data to find the likelihood probability of the fit to the data for that pair of mass values as was suggested for the KATRIN data. Another possibility would be the IceCube experiment.  IceCube has sought evidence for $\nu_\mu$ and ${\bar{\nu_\mu}}$ disappearance oscillations in 10.7 years of data.~\citep{Abbasi2025}  To date, however, IceCube has only done a limited amount of searching for multiple sterile neutrinos.~\citep{Cabrera2024}

\begin{figure}[t]
\centerline{\includegraphics[angle=0,width=1.10\linewidth]{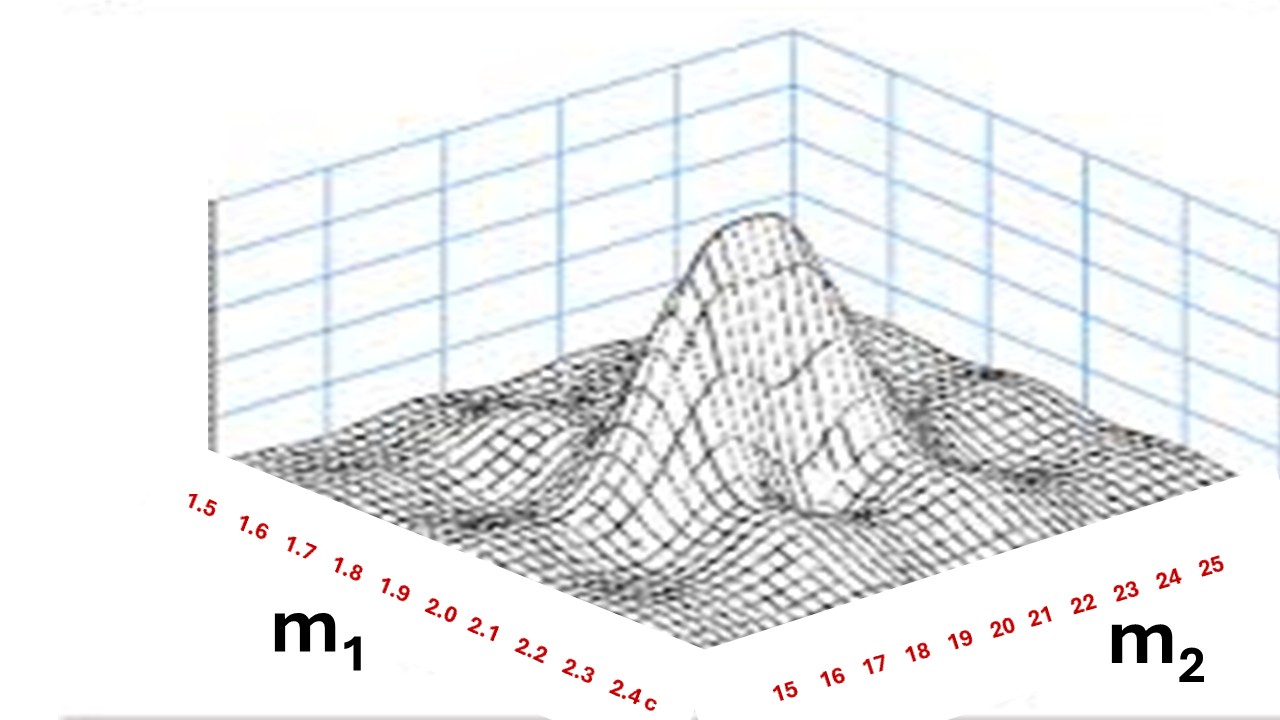}}
\caption{An imagined possible likelihood plot for the $m_1$ versus $m_2$ plane showing the maximum likelihood occurring for the $m_1$ and $m_2$ masses at the values found from the SN 1987A data. }
\end{figure}

\section{Summary and the ultimate test}
This paper has made at least four assertions that many readers may understandably consider difficult to believe -- see Table 1.  They likely will  not be accepted as true without the strongest corroborating evidence.  If KATRIN does the analysis suggested earlier in section \ref{Dog}, and it were to find evidence for the three active neutrino masses deduced from the fit to SN 1987 data, that might be sufficient. 

\begin{table}[h]
\begin{center}
\vspace{0.15in}
\begin{tabular}{ l l }
\hline
1. The LSD burst was due to SN 1987A\\
2. 3 $m_\nu^2$'s can be found from the SN 1987A data\\
3. The LSD neutrinos have $m^2<0$\\
4. The three $m^2= 2.0^2, 22.4^2$ and $-6.8\times 10^5 eV^2$\\
\hline
\end{tabular}
\end{center}
\caption{Assertions made in this paper.}
\end{table}
However, the most convincing test would be a new galactic supernova.  Given the much larger neutrino detectors today compared to those in 1987, a new galactic supernova might yield thousands of neutrinos.  Moreover, for a supernova occurring at a different distance than SN 1987A it will become completely clear if the neutrino temporal distribution reflects different travel times (and hence different masses) or instead different emission times, as most physicists believe.  This 88 year-old does not have a great chance of seeing the next one, but let's be an optimist.  Suppose supernova SN 2026(!) occurs at a distance that is a percentage p of the distance of SN 1987A.  If we were to find an LSD-like near-monochromatic burst of a thousand neutrinos occurring earlier than the main neutrino burst by a time interval of $484\times p$ minutes, there would then be little doubt that tachyons exist.  Even if the next galactic supernova does not occur in 2026, it is comforting to realize the evidence, in the form of thousands of neutrinos are now on their way to Earthly detectors.  

These neutrinos will definitely either confirm or reject the four assertions.  Moreover, given the known frequency of galactic supernovae of about two per century, we know that those neutrinos are now more than $99.99\%$ of their way here from some star in our galaxy.  No one knows which star it will be, but Betelgeuse, a massive red supergiant in Orion, is expected to end its life in a spectacular supernova explosion soon.  Soon in astronomical terms means potentially within the next 100,000 years (some say under 300 years~\citep{Saio2023}).  Go Betelgeuse!\\

\section*{Acknowledgments}
The author is grateful to Alan Chodos, Hans Thomas Janka, Francesco Vissani, Willem Elbers, Robert Ellsworth and Donald Gelman for helpful comments.   He also thanks Guido Drexlin and Diana Parno of the KATRIN Collaboration for their kindness. \\

{\bf{Conflicts of interest}\\}
The author declares no conflicts of interest.

\end{document}